\documentclass[conference]{IEEEtran}
\IEEEoverridecommandlockouts
\usepackage{cite}
\usepackage{amsmath,amssymb,amsfonts}
\usepackage{algorithmic}
\usepackage{graphicx}
\usepackage{textcomp}
\usepackage{enumitem}
\usepackage{xcolor}
\usepackage{hyperref}
\def\BibTeX{{\rm B\kern-.05em{\sc i\kern-.025em b}\kern-.08em
    T\kern-.1667em\lower.7ex\hbox{E}\kern-.125emX}}
\begin{document}

\title{Hardware-Software Co-Design for Energy-Efficient Neural Recording in Wireless Implantable BCIs}
\author{
\begin{minipage}{\textwidth}
\centering
\textbf{Junyi Wang\textsuperscript{\textdagger,1}, Hongyao Liu\textsuperscript{\textdagger,1}, Jinglong Chen\textsuperscript{2}, Liuqun Zhai\textsuperscript{1}, and Jun Huang\textsuperscript{1}}\\
\textit{\textsuperscript{1}Department of Computer Science, City University of Hong Kong}\\
\textit{\textsuperscript{2}Department of Automation, Southeast University}\\
\end{minipage}
\thanks{\textsuperscript{\textdagger}These authors contributed equally.}
}

\maketitle

\begin{abstract}
Implantable Brain-Computer Interfaces (iBCIs) are increasingly pivotal in neuroscience research and clinical neurorehabilitation. However, wireless iBCI systems are severely constrained by power dissipation limits and bandwidth bottlenecks, impeding high-fidelity, large-scale neural recording essential for advanced applications. To address these challenges, we propose a wireless iBCI headstage that implements dynamic, per-electrode ADC sampling and threshold-based spike detection. Departing from conventional application-layer compression, our system achieves efficiency at the physical acquisition layer. Specifically, an external server computes optimal, electrode-specific sampling configurations and dynamically reconfigures the front-end hardware. This approach curtails data volume before digitization, bypassing the energy overhead of computationally intensive post-processing. Extensive evaluations across diverse datasets demonstrate up to 52\% dynamic power reduction and a 3.2× decrease in hardware resource utilization, while maintaining neural decoding accuracy in both motor and visual tasks. By shifting optimization from digital compression to acquisition-layer control, this framework offers a highly practical and scalable solution for long-term, untethered neural recording.
\end{abstract}

\begin{IEEEkeywords}
iBCI, Wireless headstage, adaptive sample rate, server-driven
\end{IEEEkeywords}

\section{Introduction}
\label{sec:intro}

\begin{figure}
    \centering
    \includegraphics[width=0.99\linewidth]{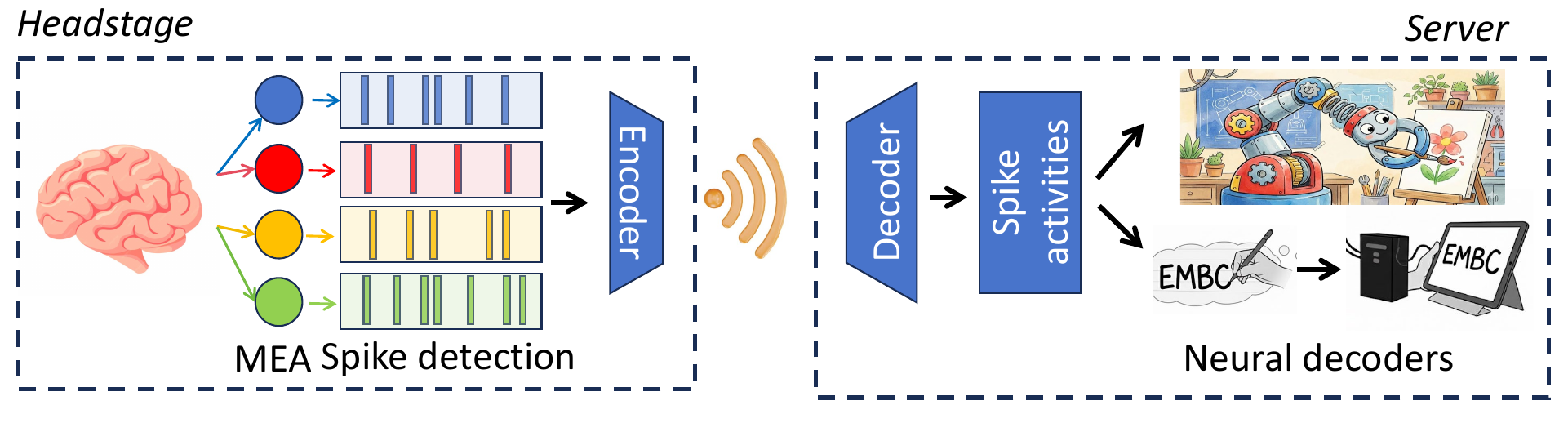}
    \vspace{-0.3in}
    \caption{\small High-level pipeline of iBCIs.}
    \vspace{-0.25in}
    \label{fig:ibci}
\end{figure}

Implantable Brain-Computer Interfaces (iBCIs) have emerged as a transformative technology for restoring functional independence in individuals with severe paralysis, such as tetraplegia and amyotrophic lateral sclerosis (ALS) \cite{simeral2021home, pun2024measuring}. Fig.~\ref{fig:ibci} illustrates the system architecture of a typical iBCI. By employing high-density microelectrode arrays (MEAs) \cite{maynard1997utah, steinmetz2021neuropixels} to record extracellular action potentials, iBCIs bypass damaged neuromuscular pathways to establish a direct communication link between the brain and external prosthetic devices. Recent clinical breakthroughs \cite{hochberg2012reach} have demonstrated the remarkable capability of iBCIs to decode complex motor intent with high degrees of freedom, enabling users to control robotic limbs \cite{wodlinger2015schwartz}, communicate with brain-to-text decoders \cite{willett2021high}, and even regain some degree of tactile feedback \cite{flesher2016intracortical}. As the field transitions from proof-of-concept laboratory demonstrations to long-term clinical applications \cite{gilja2015clinical}, the pursuit of high-fidelity, large-scale neural recording remains the cornerstone of sophisticated neural decoding.

Despite iBCIs' potential, the clinical translation of current iBCI systems is severely impeded by their reliance on percutaneous tethers. These physical connections not only introduce a persistent risk of infection at the skin interface but also anchor users to laboratory hardware, restricting mobility and natural behavior. Consequently, there is a critical imperative to transition toward fully implantable, wireless systems capable of supporting long-term home use. However, untethering the device introduces a formidable "bandwidth bottleneck". As recording densities scale to thousands of channels to enhance decoding fidelity, the aggregate raw data throughput rapidly outstrips the capacity of energy-efficient telemetry. While emerging wireless technologies such as Terahertz (THz) and Ultra-Wideband (UWB) \cite{jung2025wireless,liu2020research,liu2020rcs,liu2021switchable,liu2021analysis,zhai2026riscatter} theoretically support gigabit-level data rates, their practical utility is strictly capped by the power dissipation limits necessary to prevent thermal damage to the surrounding cortical tissue.

To mitigate the bandwidth bottleneck in high-density iBCIs, extensive research has focused on reducing data throughput via on-chip or server-assisted compression methodologies. Conventional strategies typically rely on digital-side processing, such as energy-based spike detection \cite{spikedetec2012low}, Compressive Sensing (CS) leveraging transform-domain sparsity \cite{CS2018unsupervised}, or dimensionality reduction techniques like Principal Component Analysis (PCA) and autoencoder-based feature extraction \cite{wu2018deep,fang2026inference}. While effective at reducing transmission rates, these methods often impose a paradox: effective compression requires sophisticated on-chip learning of signal characteristics, yet resource-constrained headstages lack the power and memory to support such computation without compromising feature extraction accuracy or signal fidelity. Recent server-driven architectures \cite{liu2024neuralite,liu2024neuron,liu2026sparkv} attempt to alleviate this by offloading the heavy learning burden to external processors to determine electrode-specific compression levels; however, these schemes operate largely at the application level and rely on on-chip processing after the full-resolution ADC sampling. This "digitize-then-compress" paradigm is inherently inefficient, as it expends energy digitizing clinically irrelevant noise before discarding it. Consequently, there is a critical need for approaches that shift optimization to the hardware front-end, regulating data volume directly at the acquisition source (e.g., the ADC) to maximize energy efficiency while preserving essential neural information.

To fundamentally bypass the ADC power bottleneck, we propose a server-driven, hardware-software co-design framework that shifts the compression from the digital application layer directly to the physical acquisition layer. Specifically, our wireless iBCI headstage adapts its ADC sample rate and detection thresholds on a per-electrode basis. The main contributions of this work are summarized as follows:

\begin{itemize}[leftmargin=5pt]
    \item We propose an acquisition framework that reconfigures ADC sampling on a per-electrode basis, shifting neural data reduction from post-digitization compression to source-level acquisition control.

    \item We design a lightweight server-side predictor that estimates spike detection error from spike templates under candidate sample rates and thresholds, enabling aggressive rate reduction while bounding accuracy loss.

    \item We build and validate a hardware prototype on FPGA, showing up to 52\% dynamic power reduction and 3.2$\times$ fewer hardware resources than prior digital compression approaches across motor and visual decoding tasks.
\end{itemize}

\section{Background}
\label{sec:back}

\subsection{iBCI Basics}
The iBCI framework begins with MEAs, such as Utah arrays or Neuropixels, which are implanted in the cortex to capture extracellular action potentials, commonly referred to as "spikes" \cite{pachitariu2024spike}. These spikes are transient, millisecond-scale voltage fluctuations (typically 10–500 $\mu$V) that represent the fundamental units of information transfer between neurons. The raw analog signals are first routed to a headstage located on the subject's cranium, which houses integrated analog front-ends (AFEs) for low-noise amplification, bandpass filtering, and ADC. Once digitized, the neural data is transmitted via wired tethers or wireless links to a computational server for real-time analysis. This backend processing typically involves spike detection, which isolates neural events from background noise using amplitude thresholding, and spike sorting, a critical step that assigns detected waveforms to individual neurons based on their unique morphologies. Finally, the extracted firing rates and temporal patterns are fed into a neural decoder, which employs advanced machine learning algorithms to map the brain's intent into kinematics for robotic limbs or communication interfaces. 

\begin{figure}
    \centering
    \includegraphics[width=0.99\linewidth]{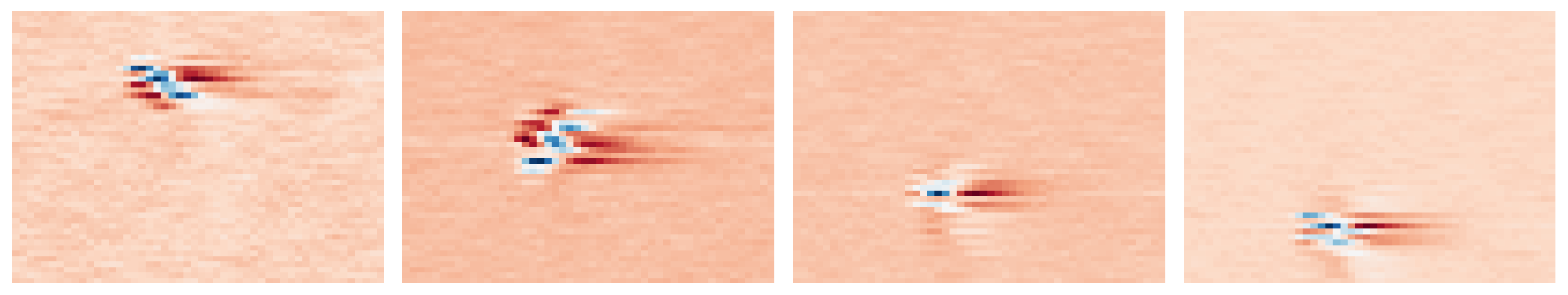}
    \vspace{-0.3in}
    \caption{\small Four example spike templates from a rat dataset \cite{rat2017fully} using Kilosort 4 \cite{pachitariu2024spike}.}
    \vspace{-0.2in}
    \label{fig:template}
\end{figure}

\subsection{Spike Sorting}
Spike sorting serves as a foundational computational process in neurophysiology, designed to identify the firing events from each neuron; however, this procedure entails substantial computational complexity, particularly when applied to large-scale datasets. Traditional spike sorting algorithms can be categorized into feature extraction \cite{wild2012performance} and template matching \cite{pachitariu2024spike}. With high-density MEAs \cite{steinmetz2021neuropixels} containing hundreds or thousands of channels, the analysis faces the critical challenge of signal superposition. In these dense recording environments, the probability of "spatiotemporal collisions", where waveforms from adjacent neurons overlap and interfere increases significantly, rendering traditional clustering methods inadequate. Consequently, the field has shifted towards template-matching algorithms, most notably exemplified by frameworks such as Kilosort \cite{pachitariu2024spike}. These approaches treat the raw data as a linear sum of individual neuronal signatures to resolve overlapping events efficiently. In this context, a "template" is defined as the high-dimensional summary of a specific neuron's electrical footprint; explicitly, it represents the average waveform profile of that neuron distributed across all relevant recording channels, capturing its unique spatiotemporal signature within the array. Fig.~\ref{fig:template} displays four spike templates from a rat 384-electrode Neuropixels dataset using Kilosort.

\subsection{Neural Decoder}
Neural decoders constitute the computational core of iBCIs, serving to translate high-dimensional neural spiking activity into actionable motor commands or communicative intent \cite{gilja2012high}. Ideally, these systems would rely on single-unit activity (SUA) isolated via spike sorting to maximize information fidelity. However, the substantial computational overhead and latency associated with real-time sorting algorithms often create a bottleneck for closed-loop control. To circumvent this, traditional decoding pipelines frequently bypass the sorting stage entirely, relying instead on threshold crossing (TC) detection \cite{spikedetec2012low}. By treating all waveform excursions beyond a noise floor as unsorted multi-unit activity (MUA), TC-based approaches significantly reduce processing complexity. Empirical studies have demonstrated that despite the loss of single-neuron resolution, such unsorted features preserve sufficient distinct information to reconstruct complex kinematics with high accuracy, thereby satisfying the strict low-latency requirements essential for fluid neuroprosthetic actuation \cite{trautmann2019accurate}.
\section{Design}

\begin{figure}
    \centering
    \includegraphics[width=0.8\linewidth]{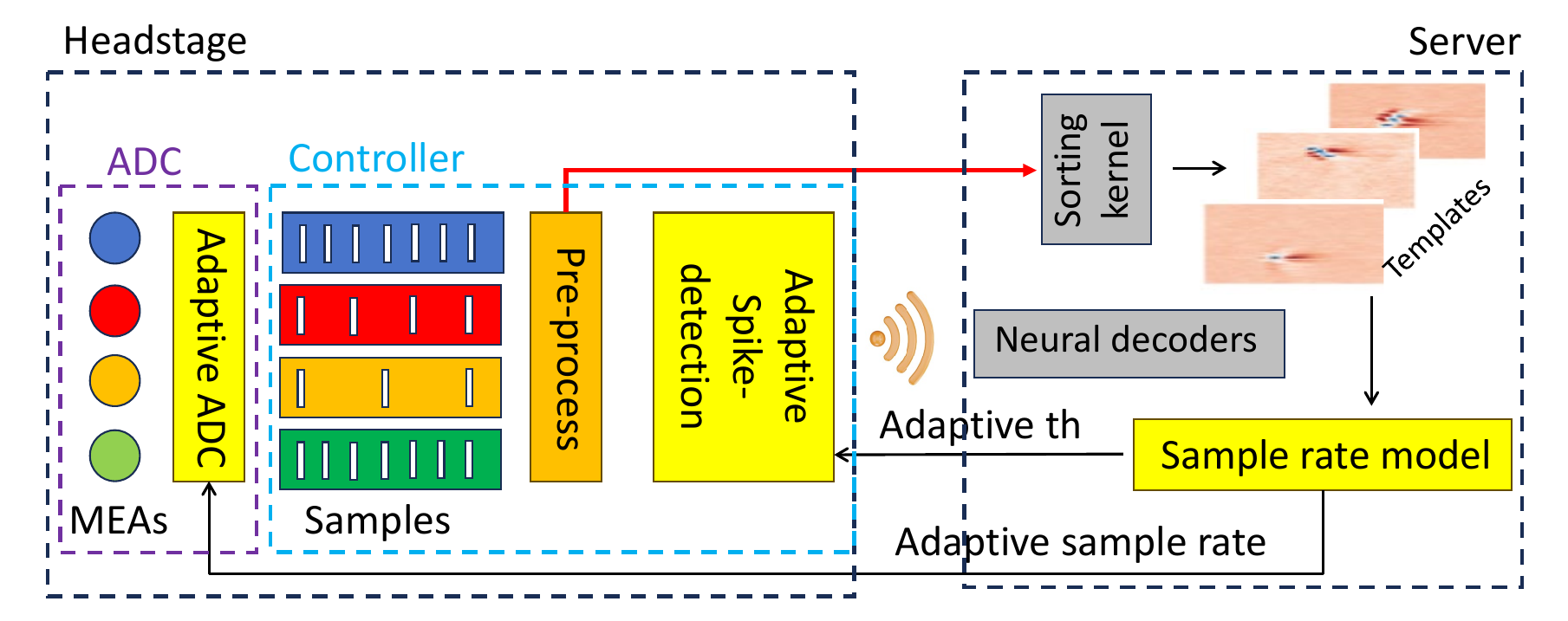}
    \vspace{-0.15in}
    \caption{\small System overview of the proposed adaptive headstage.}
    \label{fig:arch}
    \vspace{-0.2in}
\end{figure}

\subsection{Overview}
Fig.~\ref{fig:arch} shows the workflow of the proposed framework, which operates in two phases: offline calibration and online streaming. During offline calibration, the system first extracts spike templates to characterize the neural activity observed on each electrode. These templates are sent to the server, where a lightweight predictor estimates the spike detection loss associated with a candidate electrode configuration $(s_i, th_i)$, where $s_i$ and $th_i$ denote the sample rate and detection threshold of electrode $i$, respectively. Based on the predicted loss, the server selects an electrode-specific configuration that minimizes the sample rate while satisfying a user-defined error budget.

During online streaming, the optimized per-electrode schedule is delivered to the headstage through the wireless downlink and is applied directly to the ADC control logic. As a result, each electrode is physically sampled at its assigned cadence before the data enter the digital processing pipeline. The acquired samples are then filtered, whitened, and thresholded on chip before transmission. In contrast to conventional approaches that apply compression only after digitization, our method reduces data volume at the acquisition stage, thereby lowering both bandwidth and front-end energy overhead.

\begin{figure}
    \centering
    \includegraphics[width=0.8\linewidth]{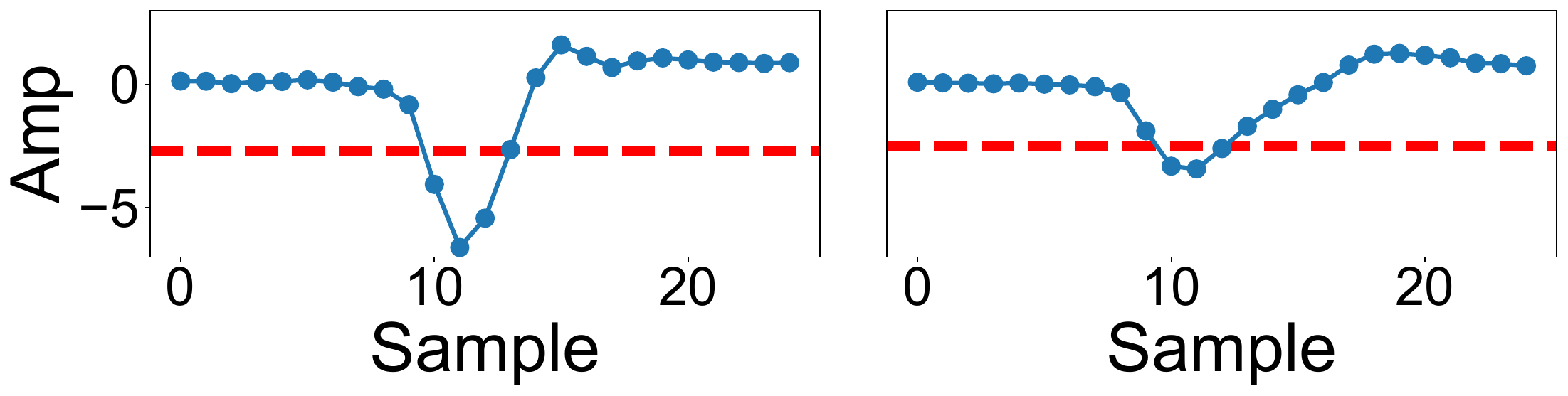}
    \vspace{-0.2in}
    \caption{\small Two example spikes detected with a threshold of -2.7$\mu V$ at the full sample rate.}
    \label{fig:sample}
    \vspace{-0.25in}
\end{figure}

\subsection{Motivation}
\textbf{Why adaptive sample rates and thresholds are feasible.}
We begin by examining spikes recorded from a single electrode. We observe that spikes with larger amplitudes remain detectable even when the sample rate is reduced, whereas weaker spikes are more sensitive to downsampling. Fig.~\ref{fig:sample} shows two representative examples under the same detection threshold. Spike A remains detectable with a downsampling factor of four, while Spike B, which has a smaller amplitude, can tolerate a factor of only three before its threshold crossing is lost. Because each electrode records a different mixture of neurons and noise, the minimum sample rate required to preserve spike detection varies substantially across electrodes. This electrode-to-electrode variability creates the opportunity for adaptive optimization.

\subsection{Predicting Detection Loss Under Adaptive Sampling}
To configure the sample rate and threshold for each electrode, we need to estimate the spike detection error induced by a candidate parameter pair $(s_i, th_i)$. This error is important because it directly affects the quality of downstream neural decoding.

\noindent
\subsubsection{Problem Formulation}
A straightforward solution would be to exhaustively evaluate all candidate configurations on raw neural recordings and compare the detection results against spike-sorting ground truth. However, repeatedly processing continuous raw traces is computationally expensive and tends to generalize poorly across recording sessions.

Instead, we use \textit{spike templates} as compact representations of neuronal activity. Templates are much lower-dimensional than raw recordings, making them efficient to process while still preserving the waveform characteristics that determine whether a spike remains detectable after downsampling. For a given electrode, we model the total detection error as two interpretable components:
\begin{enumerate}[leftmargin=10pt]
    \item \textbf{False negative rate (FNR):} the fraction of true spikes that are missed because the sample rate is too low or the threshold is too strict.
    \item \textbf{False positive rate (FPR):} the fraction of noise events that are incorrectly detected as spikes because the threshold is too permissive.
\end{enumerate}
Separating these two terms improves interpretability and allows the optimizer to distinguish between errors caused by aggressive downsampling and errors caused by over-sensitive thresholding.

\noindent
\subsubsection{Lightweight Predictor}
We implement a lightweight neural predictor that maps a candidate configuration and the electrode template to the corresponding detection error:
\begin{equation}
    f:(s_i, th_i, \mathcal{P}_i) \rightarrow (\hat{e}_{FNR}, \hat{e}_{FPR}),
\end{equation}
where $\mathcal{P}_i$ denotes the spike template waveform of electrode $i$.

\noindent\textbf{Input and output.}
The input is the concatenation of the candidate sample rate $s_i$, the detection threshold $th_i$, and the template waveform $\mathcal{P}_i$. The network outputs two scalars: the predicted FNR $\hat{e}_{FNR}$ and the predicted FPR $\hat{e}_{FPR}$.

\noindent\textbf{Architecture.}
The predictor is a compact multilayer perceptron with two hidden layers of 48 and 24 neurons, respectively, followed by ReLU activations. This design is intentionally lightweight: it is expressive enough to capture the nonlinear relationship between waveform shape, threshold, sample rate, and detection error, while remaining fast enough for practical server-side optimization.

\noindent\textbf{How the predictor is used.}
For each electrode, the server evaluates a set of candidate $(s_i, th_i)$ pairs. For each pair, the predictor returns $\hat{e}_{FNR}$ and $\hat{e}_{FPR}$, which are combined into the estimated total detection error:
\begin{equation}
    \hat{E}_{total}(s_i, th_i) = \hat{e}_{FNR} + \hat{e}_{FPR}.
\end{equation}
The server then selects the lowest sample rate whose predicted total error remains within a prescribed budget.

\subsubsection{Training and Optimization}
We train the predictor using the eMouse dataset generator~\cite{pachitariu2024spike}, a standard tool for spike-sorting validation. To generate one training sample, we insert a spike template into a 10~s segment of Gaussian white noise at random times, apply a candidate downsampling factor and detection threshold, and then compare the detector output with the known spike ground truth to obtain the corresponding FNR and FPR. In total, we generate 2,000 samples and split them into 80\% training data and 20\% testing data.

The model is trained with Stochastic Gradient Descent (SGD) to minimize the Mean Squared Error (MSE) between predicted and ground-truth error rates. We also adopt early stopping to avoid overfitting and improve generalization across signal conditions.

\noindent\textbf{Configuration strategy.}
For each electrode, we choose the minimum sample rate that satisfies a maximum allowable detection error $\epsilon$:
\begin{equation}
    \min s_i \quad \text{s.t.} \quad E_{total}(s_i, th_i) \leq \epsilon.
\end{equation}
The value of $\epsilon$ depends on the robustness of the downstream decoder. For conventional decoders, we use $\epsilon = 5\%$. For more robust decoders that can tolerate moderate front-end errors, the constraint can be relaxed to $\epsilon = 10\%$ to enable more aggressive compression and larger power savings.

\begin{figure}[t]
    \centering
    \includegraphics[width=0.8\linewidth]{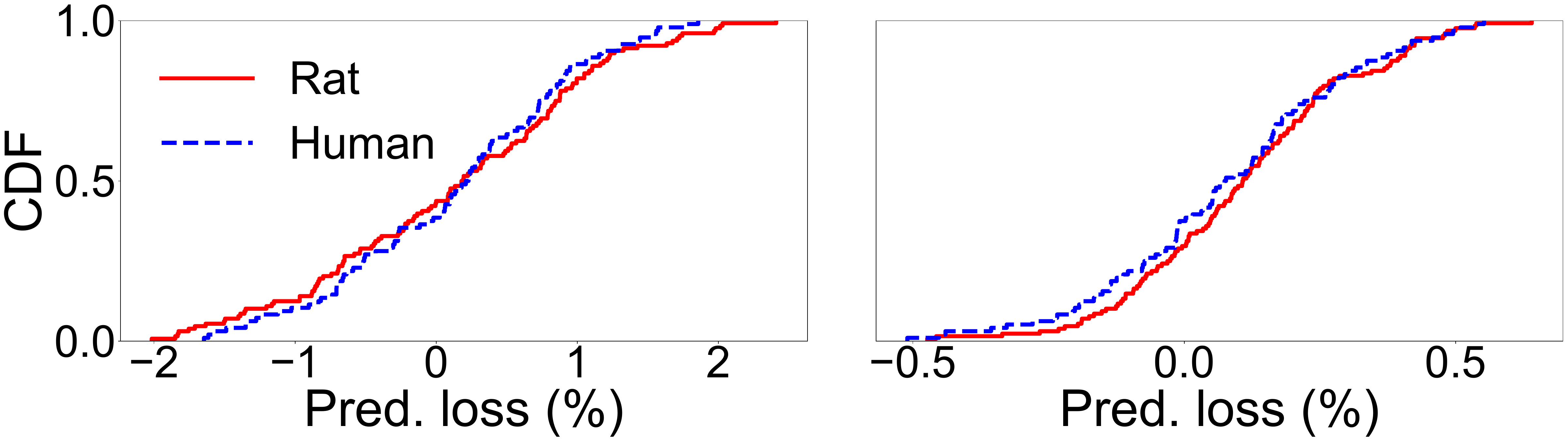}
    \vspace{-0.1in}
    \caption{\small Prediction error of FNR (left) and FPR (right) on rat and human datasets.}
    \label{fig:pred}
    \vspace{-0.15in}
\end{figure}

We evaluate the predictor on a rat dataset~\cite{rat2017fully} and a human cortex dataset~\cite{human2022large}, both recorded with Neuropixels MEAs~\cite{steinmetz2021neuropixels}. Fig.~\ref{fig:pred} shows the prediction error distribution across electrodes. The predictor achieves errors below 2\% for FNR and below 0.6\% for FPR. Importantly, this model is trained only once offline and runs on the server, so it introduces no additional area or power overhead to the headstage. The inference latency is 4.3~ms, and the optimization of $\{s_i, th_i\}$ converges within 46~s on an Intel Core i7-13500KF CPU, which is sufficiently fast to track template drift that typically occurs over tens of minutes~\cite{liu2024neuralite}.

\subsection{ADC-Level Sample Rate Adjustment}
Conventional iBCI streaming architectures usually perform data reduction only after digitization. While simple to implement, this strategy still pays the cost of ADC conversion, buffering, and digital processing for samples that are later discarded. Our system instead enforces adaptive sampling directly at the ADC/FPGA interface, so unnecessary samples are never acquired in the first place.

Let $\mathbf{S} = \{s_1, s_2, \dots, s_N\}$ denote the target sample-rate vector, where $s_i$ is the desired sample rate of electrode $i$. The ADC interface is driven by a master clock $f_{clk}$ and executes a periodic command schedule. If each execution round contains $N_t$ clock cycles, among which $N_s$ cycles correspond to active sampling commands, then the maximum achievable sample rate is
\begin{equation}
    R_{max} = \frac{N_s \cdot f_{clk}}{N_t}.
\end{equation}

To realize a lower sample rate, each electrode is assigned an integer downsampling factor $x_{d,i}$ from a supported set of hardware factors. The FPGA then samples electrode $i$ once every $x_{d,i}$ scheduling rounds. The implemented rate is therefore
\begin{equation}
    \hat{s}_i = \frac{R_{max}}{x_{d,i}}.
\end{equation}
To ensure that the realized hardware rate is the closest feasible rate that still meets the target, we choose
\begin{equation}
    x_{d,i} = \max \left\{ x \in \mathcal{X} \; \middle| \; \frac{R_{max}}{x} \ge s_i \right\},
\end{equation}
where $\mathcal{X}$ is the set of supported downsampling factors.

Operationally, the scheduler is extremely simple. Let $r$ denote the current execution round. For electrode $i$, the FPGA asserts its sampling command only when
\begin{equation}
    r \bmod x_{d,i} = 0.
\end{equation}
Otherwise, the corresponding sampling opportunity is skipped. For example, an electrode with $x_{d,i}=1$ is sampled every round, an electrode with $x_{d,i}=2$ is sampled every other round, and an electrode with $x_{d,i}=4$ is sampled once every four rounds. This periodic schedule can be implemented using only simple counters and clock-division logic, without digital decimation filters or extra buffering.

\section{Implementation}

\begin{figure}
    \centering
    \includegraphics[width=0.95\linewidth]{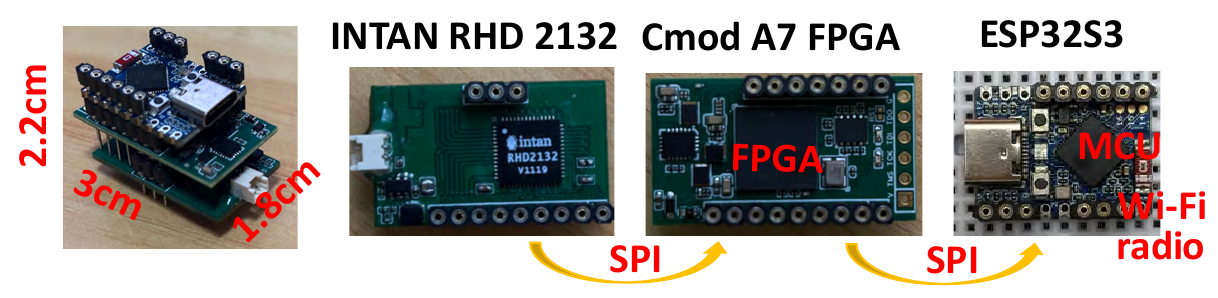}
    \vspace{-0.1in}
    \caption{\small Architecture of the prototype.}
    \vspace{-0.2in}
    \label{fig:photo}
\end{figure}

\noindent
\textbf{Headstage Architecture.}
Fig.~\ref{fig:photo} shows our wireless headstage prototype ($<$173~mW). It comprises an Intan RHD2132 analog front-end, a Cmod A7 FPGA, and an ESP32-S3 wireless transceiver, operating in a server-guided closed loop.

Each adaptation interval, the server computes and transmits a per-electrode configuration vector $\mathbf{C} = \{(s_i, th_i)\}_{i=1}^{N}$, specifying the sample rate $s_i$ and detection threshold $th_i$ for each electrode. The FPGA drives the Intan front-end via SPI according to the configured schedule, enforcing rate reduction directly at the acquisition stage. Acquired samples pass through an on-chip pipeline of band-pass filtering, whitening, and threshold-based spike detection. Only detected spike events are packetized and wirelessly transmitted to the server via the ESP32-S3, avoiding raw broadband transmission. This design offloads intensive computation to the server while keeping the headstage limited to schedule execution, lightweight signal conditioning, and low-bandwidth event transmission.

\section{Evaluation}

\subsection{Methodology} 

\noindent
\textbf{Datasets.} We evaluate the headstage prototype on five iBCI datasets. Three raw neural recordings from macaque \cite{macaque2018massively}, rat \cite{rat2017fully}, and human \cite{human2022large} subjects, acquired with Utah and Neuropixels MEAs, are used to assess spike detection performance. In addition, we use a visual and a motor decoder from CEBRA \cite{schneider2023learnable} to assess the impact of downsampling on downstream tasks.

\noindent
\textbf{Baselines.} We benchmark our approach against two neural signal compression algorithms implemented on the FPGA: Discrete Cosine Transform (DCT) \cite{aprile2017dct} and Compressive Sensing (CS) \cite{CS2018unsupervised}. Both schemes represent application-layer implementations that process fully digitized neural signals.

\noindent \textbf{Evaluation Metrics.}
We assess system performance across two dimensions:
(1) \textit{Signal Fidelity:} We quantify the \textit{Compression Ratio (CR)} and the \textit{Spike Detection Error (SDE)} (comprising false negatives and false positives) to measure data efficiency and integrity. (2) \textit{Hardware Efficiency:} We profile the power consumption (mW) and FPGA resource utilization (LUTs/Flip-Flops) to determine the implementation overhead.

\noindent
\textbf{Experimental Setup.}
We conduct a comprehensive data-driven evaluation. Datasets recorded with 100- and 384-electrode MEAs were partitioned into 32-channel subsets to match the acquisition capacity of the Intan system.

\subsubsection{Electrode-Level Ground Truth Construction}
Spike sorting yields neuron-level labels, whereas our ADC controller operates at electrode granularity. To bridge this mismatch, we project neuron-level spike trains onto the electrode array. Let $y_n(t) \in \{0,1\}$ denote the spike train of neuron $n$, and let $\mathcal{E}_n$ be its spatial footprint, i.e., the set of electrodes on which neuron $n$'s waveform exceeds the noise floor. The electrode-level ground truth is
\begin{equation}
    G_e(t) = \bigvee_{n:\,e \in \mathcal{E}_n} y_n(t),
\end{equation}
so that $G_e(t)=1$ whenever at least one neuron visible on electrode $e$ fires at time $t$. Simultaneous spikes from multiple neurons on the same electrode are treated as a single detection event, consistent with the per-electrode threshold detector.

\subsubsection{ADC Simulation}
Since raw datasets are already digitized, we employ a pipeline to mimic the behavior of the Intan RHD2132 chip. We first reconstruct the continuous analog waveform from the discrete dataset using band-limited interpolation. Subsequently, we simulate the hardware ADC and amplifier chain by re-digitizing this waveform using the specific adaptive sample rate vectors derived by our scheduler. This ensures the digital signal faithfully reflects the artifacts introduced by variable-rate hardware sampling.

\subsubsection{Stochastic Error Injection for Decoders}
For the CEBRA decoding benchmarks, raw neural traces are unavailable, precluding direct resampling. Instead, we assess robustness by injecting the empirical errors observed in the rat dataset. We perturb the ground truth spike sequences $\mathbf{y}$ to generate corrupted traces $\hat{\mathbf{y}}$ as follows:
(1) \textbf{Missed Spikes (FNR):} True spikes are removed (set to 0) with a probability equal to the empirical FNR of the target electrode observed in the rat dataset. (2) \textbf{Spurious Spikes (FPR):} Spurious spikes are inserted (0 flipped to 1) at random idle timesteps with a probability derived from the empirical FPR of the corresponding electrode in the rat dataset.

\subsection{Compression efficiency}
\begin{figure}[t]
    \centering
    \includegraphics[width=0.47\linewidth]{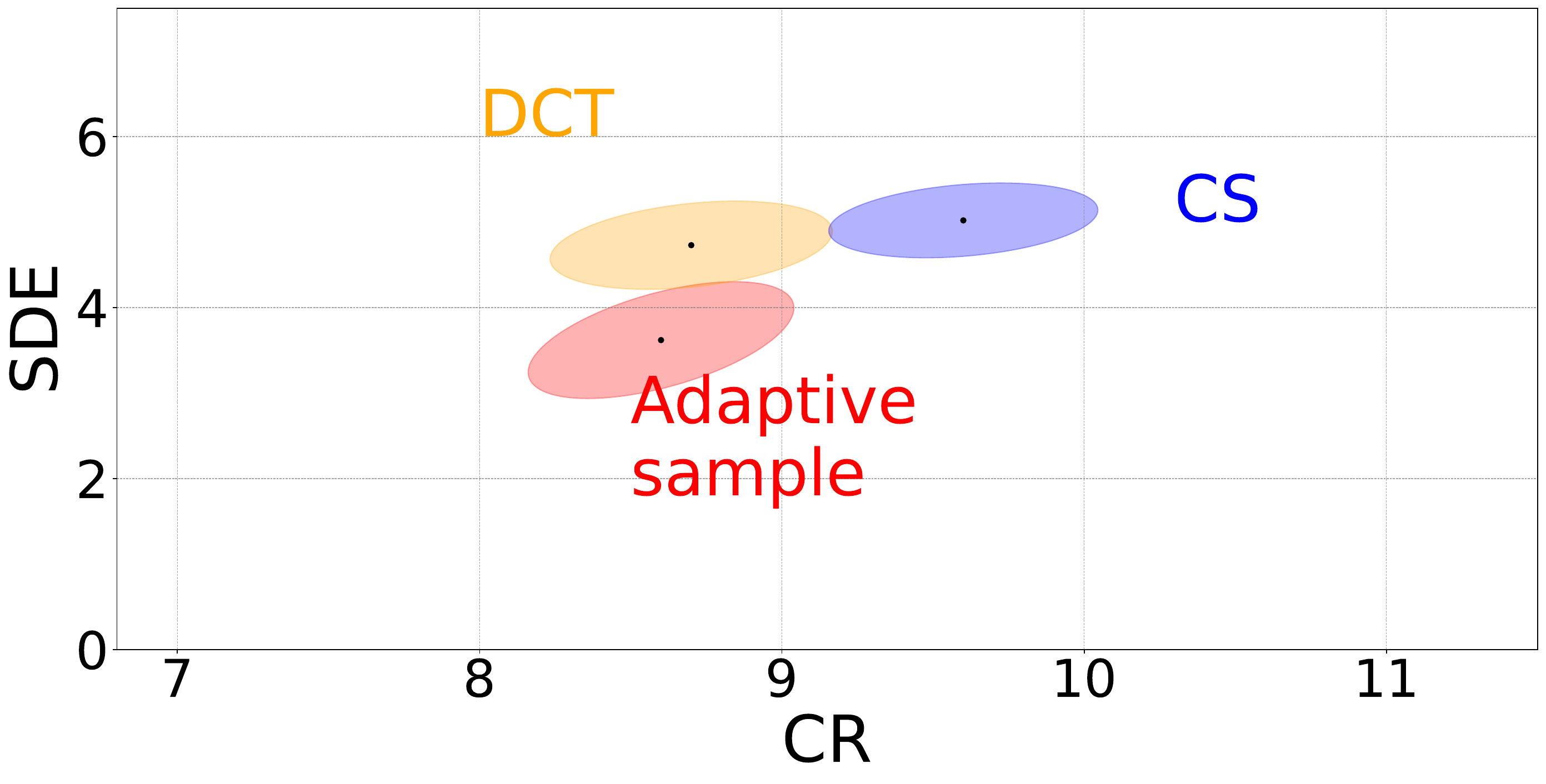}
    \includegraphics[width=0.47\linewidth]{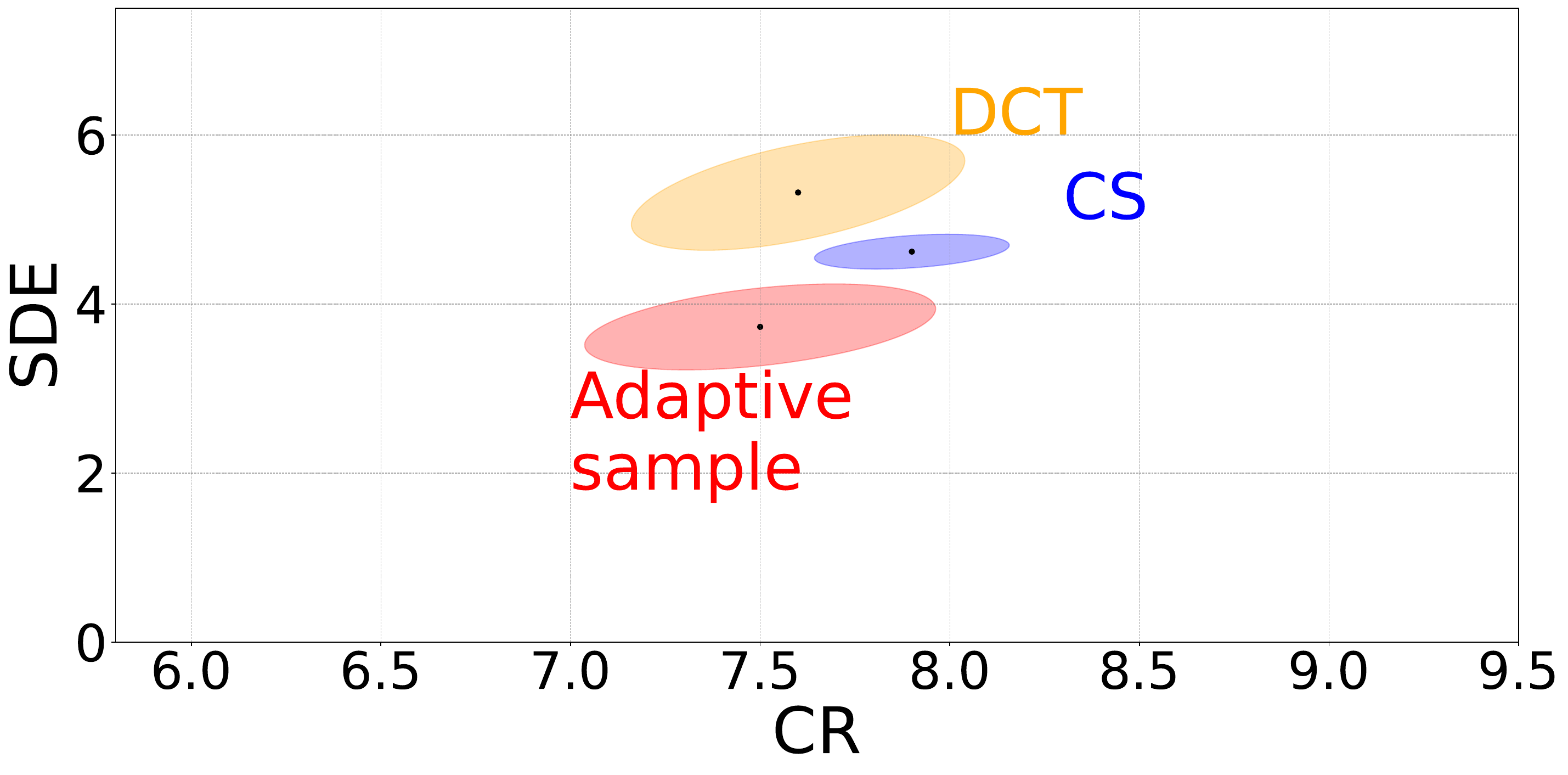}
    \vspace{-0.15in}
    \caption{\small CR versus SDE across macaque and human datasets using three neural signal streaming schemes.}
    \vspace{-0.15in}
    \label{fig:comp}
\end{figure}

To assess the compression efficiency, we compare the compression ratio of our proposed adaptive sample rate mechanism with the other two baselines on macaque and human datasets. As shown in Fig.~\ref{fig:comp}, our ADC-level adaptive sampling achieves a compression ratio comparable to DCT and CS while reducing the spike detection error by approximately 2\%. Although the absolute compression ratio of sample rate adjustment alone does not surpass these two transform-based algorithms, the electrode-adaptive allocation consistently yields lower detection error than uniformly compressing the neural signal. In addition, the adaptive spike detection threshold provides more accurate spike identification at the same compression ratio compared with uniform or noise-based thresholding.

\subsection{Power efficiency}
\begin{figure}[t]
    \centering
    \includegraphics[width=0.99\linewidth]{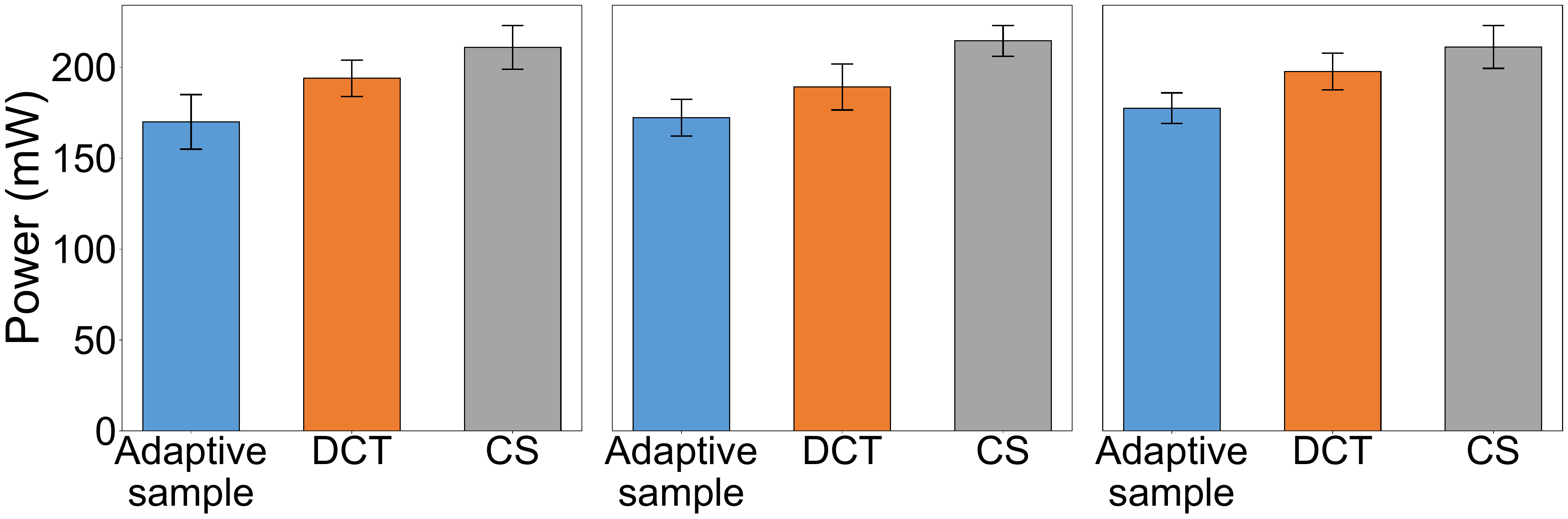}
    \vspace{-0.25in}
    \caption{\small Comparison of power consumption between adaptive sample and other two baselines across macaque, rat and human datasets.}
    \vspace{-0.2in}
    \label{fig:power}
\end{figure}

To evaluate the power efficiency of the proposed adaptive ADC sampling mechanism, we measured the power consumption during the streaming process using a power meter. Fig.~\ref{fig:power} compares the power consumption across three datasets, demonstrating that our adaptive sampling method achieves the highest power efficiency. Specifically, the adaptive ADC sampling reduces power consumption by 24--35 mW compared to the DCT implementation on the FPGA layer. Furthermore, compared to CS, the adaptive approach yields even greater savings of 29--40 mW, as CS is computationally more intensive than DCT. Since the detected spike activities transmitted via Wi-Fi constitute a small data payload, the radio operates at a low duty cycle. Consequently, the majority of power is consumed by the Intan ADC chip and FPGA processing. Direct configuration of the ADC sample rate proves to be more power-efficient because it primarily reduces the sampling operations of the ADC chip itself. Additionally, the FPGA layer only needs to process significant samples, resulting in reduced buffer usage and lower computational overhead.

\subsection{Resource utilization efficiency} 
\begin{figure}[t]
    \centering
    \includegraphics[width=0.85\linewidth]{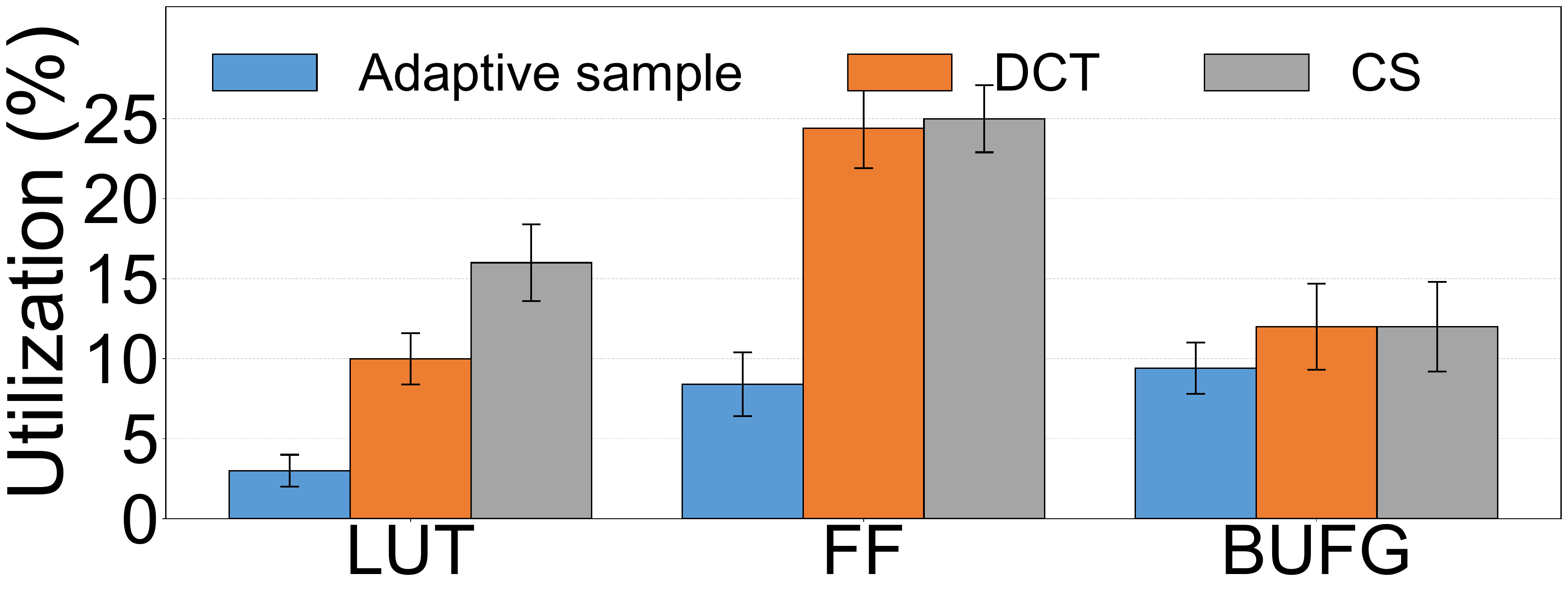}
    \vspace{-0.2in}
    \caption{\small FPGA resource utilization percentage among three neural signal streaming schemes in rat dataset.}
    \vspace{-0.2in}
    \label{fig:proc}
\end{figure}

To evaluate hardware efficiency, we compared the resource utilization on a Cmod A7 FPGA as reported by Vivado 2025. Fig.~\ref{fig:proc} illustrates the consumption of the three most critical on-chip resources: Look-Up Tables (LUT), Flip-Flops (FF), and Global Clock Buffers (BUFG). The results demonstrate that our adaptive ADC sampling method significantly alleviates the digital processing burden. Specifically, it reduces LUT utilization by $2.8\times$ and $3.2\times$ compared to DCT and CS, respectively, confirming its streamlined logic architecture. Furthermore, the adaptive approach reduces FF utilization by roughly $3\times$ and exhibits lower complexity in the clock domain, as evidenced by the reduced BUFG usage.

\subsection{Impact on Neural Decoders}
\begin{figure}[t]
    \centering
    \includegraphics[width=0.99\linewidth]{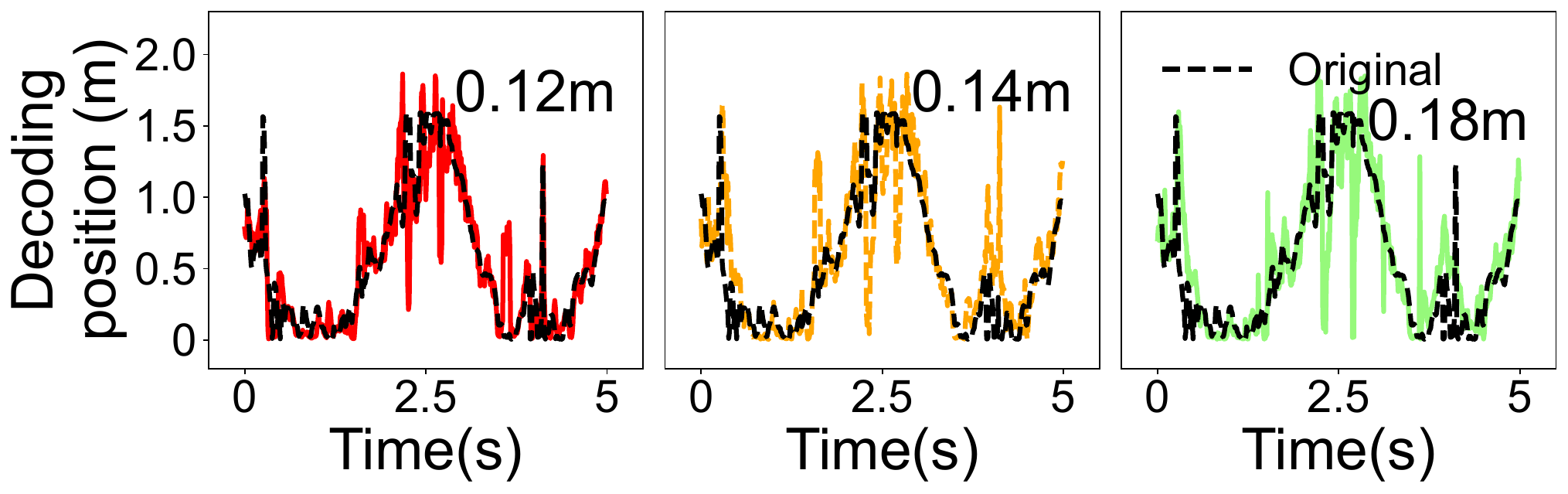}
    \vspace{-0.3in}
    \caption{\small Impact of adaptive sample, CS, and DCT on the motor decoder from CEBRA.}
    \vspace{-0.15in}
    \label{fig:motor}
\end{figure}

We further evaluated the downstream impact of our adaptive sampling wireless headstage on both visual and motor decoding tasks. The motor decoder translates neural spiking activity into the estimated trajectory of the rat. As illustrated in Fig.~\ref{fig:motor}, the proposed adaptive ADC sampling mechanism achieves the highest accuracy, with a median trajectory prediction error of 0.12~m. This represents a reduction in error of 0.02~m and 0.06~m compared to CS and DCT, respectively. The superior performance of our headstage is directly linked to its computational efficiency. While all methods maintain comparable compression ratios, the adaptive sampling approach imposes significantly lower signal processing overhead. In contrast, the higher computational complexity of DCT and CS may introduce processing latency that risks buffer overflows and the loss of neural signal frames during high-activity periods. By minimizing these processing delays, our headstage scales effectively with data volume and MEA size, making it particularly suitable for closed-loop applications with strict latency requirements.

\begin{figure}[t]
    \centering
    \includegraphics[width=0.8\linewidth]{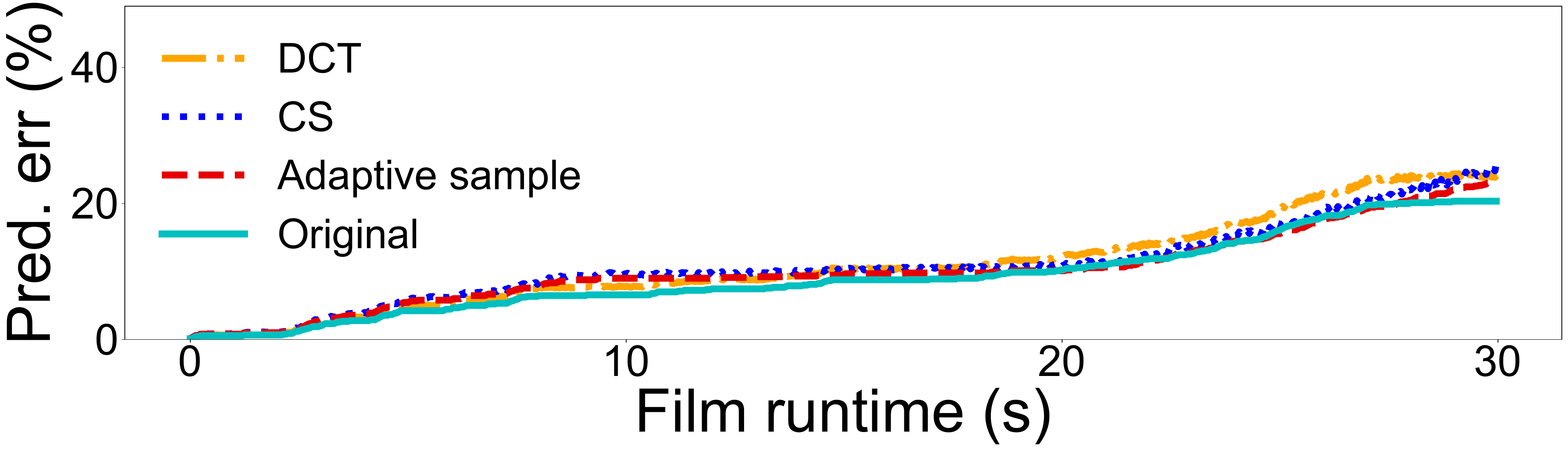}
    \vspace{-0.1in}
    \caption{\small Impact of adaptive sample, CS, and DCT on the video decoder from CEBRA.}
    \vspace{-0.2in}
    \label{fig:video}
\end{figure}

The visual decoder predicts the ID of the frame currently being viewed by the rat, using a stimulus film running at 30~fps. Fig.~\ref{fig:video} depicts the accumulated prediction error as the video playback time progresses from 0 to 30 seconds. We observe that all three streaming schemes converge to a similar decoding accuracy by the end of the session. However, performance parity alone does not capture the full system impact. When power consumption and processing efficiency are taken into account, our adaptive approach proves superior. Unlike CS and DCT, which require substantial hardware resources to maintain this accuracy, our headstage with ADC sample rate control achieves these results with minimal power overhead. This indicates that our design is ready for deployment in real-world experiments without modification and is a practical solution for future iBCI MEAs that demand support for large-scale data and advanced neural decoders.

\section{Discussion and Limitations}

\noindent
\textbf{Flexible sample rate modulation.} A primary limitation of the current framework is that the sample rate modulation is restricted to integer downsampling factors (e.g., $\times 2, \times 3$). This coarse granularity prevents the system from precisely matching the target sample rate to the signal's sparsity. Consequently, the allocated sample rate is often forced to the next available integer tier, leading to residual bandwidth inefficiency. Future iterations of this work will aim to implement fractional downsampling schemes (e.g., $3/4$) to achieve higher-resolution frequency adjustment and minimize over-sampling.

\noindent
\textbf{ASIC Implementation.}
While our current headstage prototype leverages the Cmod A7 FPGA for rapid development and algorithmic validation, it is constrained by the inherent power overhead of reconfigurable logic. To address this, we plan to transition the design to an Application-Specific Integrated Circuit (ASIC). By migrating from general-purpose FPGA fabric to dedicated silicon, we project a power reduction of $6\text{--}9\times$, bringing the total consumption to under 30~mW. This transition will not only preserve the high compression and processing efficiency of our adaptive sampling mechanism but also significantly reduce the physical footprint and weight, making it ideal for chronic experiments on small, freely moving animals.

\noindent
\textbf{Server dependency and scalability.} Our framework relies on a server to compute the per-electrode configuration. In practice, this optimization runs only during periodic recalibration intervals (on the order of minutes), not on every sample. Between recalibrations, the headstage operates autonomously using the last received schedule. If the wireless downlink is temporarily interrupted, the headstage continues streaming with the most recent configuration, gracefully degrading rather than failing. Regarding scalability, the per-electrode optimization is inherently independent across channels, allowing the server-side computation to parallelize linearly with channel count. The on-chip scheduling logic (modulo counters) also scales with negligible area overhead per electrode, making the architecture well-suited for next-generation MEAs with thousands of channels.

\noindent \textbf{Future Work: Acute \textit{In-Vivo} recording.} To bridge the gap between simulation and clinical application, we plan to transition from offline emulation to real-time \textit{in vivo} experiments. In collaboration with the Department of Neuroscience at our institution, we will conduct acute implantation experiments on anesthetized mice. The choice of an acute, anesthetized model represents a strategic intermediate step before chronic implantation in behaving animals. These experiments will validate three key aspects: (1) thermal safety of the headstage under continuous operation near cortical tissue, (2) real-time spike detection fidelity compared with the offline emulation results reported in this work, and (3) the stability of the adaptive configuration under physiological noise conditions that are absent in curated datasets.

\section{Conclusion}
In this work, we propose a wireless iBCI headstage featuring lightweight, adaptive control of ADC sample rates and spike detection thresholds. Unlike conventional signal compression systems that rely on complex processing in the application layer, our approach directly reconfigures downsampling at the ADC layer. This server-driven design represents the first framework to compress data at the true source of the stream. Extensive experiments demonstrate that our headstage achieves significant savings in power and FPGA resource utilization while maintaining high compression efficiency and decoding accuracy.

\section*{Acknowledgments}

This work is supported in part by the Research Grants Council (RGC) of Hong Kong under General Research Fund under Grant 11220225. 

\bibliographystyle{IEEEtran}
\bibliography{Reference}

\end{document}